\documentclass{JHEP3} 

\JHEPspecialurl{http://jhep.sissa.it/JOURNAL/JHEP3.tar.gz}

\usepackage{epsfig,multicol,bbm}
\usepackage[verbose]{cite}

\newcommand\fverb{\setbox\fverbbox=\hbox\bgroup\verb}
\newcommand\fverbdo{\egroup\medskip\noindent%
			\fbox{\unhbox\fverbbox}\ }
\newcommand\fverbit{\egroup\item[\fbox{\unhbox\fverbbox}]}
\newbox\fverbbox

\usepackage{braket}
\newcommand{\Tr}{\mathrm{Tr}}
\usepackage{amsmath}	

\title{Regularization of identity based solution in string field theory}

\author{Syoji Zeze\\
	Yokote Seiryo Gakuin High School,\\ 
	147-1 Maeda, Osawa, Yokote,\\
	JAPAN 013-0041
	E-mail: \email{ztaro21@gmail.com}
}

\preprint{\arXivid{XXXX.XXXX}}	

\abstract{
We demonstrate that an Erler-Schnabl type solution in cubic 
string field theory can be naturally interpreted as a gauge invariant 
regularization of an identity based solution. 
We consider a solution which 
interpolates between an identity based solution and ordinary 
Erler-Schnabl one.   Two gauge invariant quantities, 
the classical action and the closed string tadpole, are evaluated for
finite value of the gauge parameter.   It is explicitly checked that 
both of them are independent of the gauge parameter.  
}

\begin{document}

\section{Introduction}

Identity based solutions, which are constructed upon the identity string
field, have been mysterious objects in string field theory.
Typically a solution takes form of
\begin{equation}
 \Psi = C I,
\end{equation} 
where $C$ is certain linear combination of 
ghost number one operators and $I$ is the identity string field, 
a surface state which represents
an open string world sheet of vanishing width.  
Such solutions have been considered since
early days of string field theory
\cite{Horowitz:1986dta,Horowitz:1987kz,Horowitz:1987yz}.  
More elaborated versions have been investigated \cite{%
Takahashi:2001pp,Kishimoto:2001de,
Takahashi:2002ez,Kishimoto:2002xi,
Takahashi:2003xe,Takahashi:2003kq,
Zeze:2004yh,Zeze:2005qt,
Igarashi:2005wh,Igarashi:2005sd,
Kishimoto:2009nd, Kishimoto:2009hc
} to describe
tachyon condensation and marginal deformation.  
Even after
Schnabl's discovery of the analytic solution \cite{Schnabl:2005gv}, 
it had  been recognized that an identity based string field
is useful to construct regular solutions 
\cite{Erler:2006hw, Erler:2006ww, Schnabl:2007az, Kishimoto:2007bb}.
However, even though much efforts have been done in past, 
identity based solutions have not yet been widely 
accepted as a regular solution.
A major problem is indefiniteness 
of physical quantities which 
originates from the inner product between identity based string
fields.   Naive evaluation of the
classical action in terms of CFT method tends to be
indefinite since it corresponds to a correlator 
on vanishing strip.  Various attempts of regularization
in terms of a strip of infinitesimal width 
still seem to fail to give definite value of the classical action
\cite{Kishimoto:2001de, Arroyo:2010fq, Zeze:2010jv}. 

In this paper, we consider one-parameter family of classical solutions
in open string field theory given by
\begin{equation}
 \Psi_{\lambda} = U_{\lambda} Q_B U_{\lambda}^{-1}
+U_{\lambda} \Psi_{I} U_{\lambda}^{-1},
\end{equation}
where $\Psi_{I}$ is an identity based solution and $U_{\lambda}=1+ \lambda c
B K$ is an element of gauge group \cite{Arroyo:2010fq}.  Since
$U_{\lambda}$ approaches to identity as $\lambda$ goes to zero,  it
is clear that the above string field naturally definies gauge invariant 
regularization which interpolates between an identity based solution
and non identity one.  We will see that suitable choice of $\Psi_{I}$
yields an Erler-Schnabl type solution.  

Rest of the paper is devoted to evaluation of two 
gauge invariant quantities, the classical action 
and the closed string tadpole, following to a method used in
\cite{Erler:2009uj, Zeze:2010jv}.  It will be checked that 
both of them are independent of the gauge parameter $\lambda$.

\section{Regularization of an identity based solution}

To begin with, let us briefly review the notation used in 
\cite{Okawa:2006vm, Erler:2006ww, Erler:2006hw, Erler:2009uj} 
 which will be extensively used in this paper.
The building block of our solution is elements of `$K B c$ subalgebra'
which is defined by
\begin{equation}
 \{B, c\} =1, \quad [B, K] =0,  \quad \{B, B\} =0, \quad \{c, c\}=0.
\end{equation}
The action of the BRST charge on these elements is
\begin{equation}
 Q_{B} c = c K c,  \quad  Q_{B} B = K, \quad Q_{B} K =0.
\end{equation}
In this notation, the star multiplication between elements is understood.
The BPZ inner product of elements is denoted as `trace'.  

Next, let us regularize a simple identity based 
solution appeared in \cite{Erler:2009uj, Arroyo:2010fq, Schnabl:2010tb}.
A clue to regularization is Arroyo's observation made in
\cite{Arroyo:2010fq}, where it was shown that the identity based
solution
\begin{equation}
 \Psi_{A} = c (1-K) \label{223823_17Jun10}
\end{equation}
is gauge equivalent to the Erler-Schnabl solution.
\begin{align}
 \Psi_{ES} & = c (1-K) B c \frac{1}{1+K} \notag \\
  & = U c (1-K) U^{-1} + U Q_{B} U^{-1},\label{225328_27Jun10} 
\end{align}
where $U$ is a gauge transformation given by
\begin{equation}
 U = 1+ c B K, \quad U^{-1} =1 - c B K \frac{1}{1+K}. \label{gauge}
\end{equation}
Here we consider a slight modification of (\ref{gauge})
in which a real parameter is inserted in front of the $ c B K$ piece
in the gauge transformation. 
\begin{equation}
 U_{\lambda } = 1+ \lambda c B K, \quad 
U_{\lambda}^{-1} =1 - \lambda c B K \frac{1}{1+\lambda K}.
\end{equation}
Then, one-parameter family of solutions is obtained by performing
the above gauge transformation on the identity bases solution  (\ref{223823_17Jun10})
\footnote{A real solution can be obtained by putting
$\sqrt{(1+\lambda K)/(1+(\lambda-1)K )} $ on both sides of 
the solution. The author thank H.~Isono for discussion.
}.
\begin{align}
 \Psi_{\lambda} & \equiv U_{\lambda} Q_{B} U_{\lambda}^{-1} + 
U_{\lambda } c(1-K) U_{\lambda}^{-1} \notag \\
& = c (1+ \lambda K) B c \frac{1+ (\lambda -1) K }{1+ \lambda K}.\label{225510_17Jun10}
\end{align}
A check of equation of motion for the above solution is straightforward.
It should be noticed that the solution resembles the non-real form of
Erler-Schnabl's
solution of \cite{Erler:2009uj}.  Then we are interested in how general
 such solution is.   Fortunately, a wider class of solutions
is already considered in \cite{Erler:2009uj} and further explored in 
\cite{Schnabl:2010tb}.  Here we focus on a class of solutions of the 
form
\begin{equation}
 \Psi = c F B c\,  G,
\end{equation}
where $F$ and $G$ are functions of $K$.  Equation of motion for this
field becomes
\begin{equation}
 c \left\{ (F G + K) c F
- F c (F G +K)
\right\}B c G =0.\label{235831_4Aug10}
\end{equation}
There are two nontrivial solutions of  (\ref{235831_4Aug10}),
\begin{equation}
 F G + K  = F, \quad F G + K  =0,
\end{equation}
and corresponding string fields
\begin{equation}
 \Psi_{1} = c F B c \left(1- \frac{K}{F}\right), \quad
\Psi_{2} = - c F B c \frac{K}{F}.\label{003651_5Aug10} 
\end{equation}
It can be easily seen that the solution (\ref{225510_17Jun10})
corresponds to  $\Psi_1$ with $F= 1+ \lambda K$.

Next, let us evaluate classical action in CFT method. 
As for a classical solution of the equation of motion, we only need to
evaluate
\begin{equation}
E = \frac{1}{6} \Tr [\Psi_{\lambda} Q_{B} \Psi_\lambda   ].  
\end{equation}
We evaluate this quantity following to \cite{Erler:2009uj,
 Zeze:2010jv}.  As similar to the case of \cite{Erler:2009uj},
the second term of (\ref{225510_17Jun10}) which begins from $c \lambda K$
does not contribute to $E$.  Introduction of Schwinger
parameters leads
\begin{multline}
 \Tr [\Psi_\lambda Q_{B} \Psi_\lambda   ] =
\int_{0}^{\infty} \int_{0}^{\infty} 
d t_1 d t_2 e^{-t_1 -t_2}\\
\times \Tr [
c (1+(\lambda -1) K) \Omega^{\lambda t_1}
c K c (1+(\lambda -1) K)  \Omega^{\lambda t_2}
]. \label{231449_17Jun10}
 \end{multline}
All terms in (\ref{231449_17Jun10}) can be obtained by
differentiating the basic trace $h(\lambda t_1, \lambda t_2 )$ with
respect to $t_1$ or $t_2$. Here, the 
basic trace $h(s,t )=\Tr [  c \Omega^{s} c K c \Omega^{t}     ]$ is
already known to be
\begin{equation}
 h(s,t ) = -\left(\frac{s+t}{\pi}\right)^2 \sin^2 \left(
\frac{\pi s}{s+t}
\right).
\end{equation}
The trace in the right hand side of  (\ref{231449_17Jun10}) 
is can be written as
\begin{equation}
 (\text{trace}) =
\left(1
  +\frac{1-\lambda}{\lambda} \partial_{t_1}\right)
\left(1
  +\frac{1-\lambda}{\lambda} \partial_{t_2}\right)
h(\lambda t_1, \lambda t_2 ).
\label{234157_17Jun10}
\end{equation}
Remaining process is completely parallel to that of
\cite{Erler:2009uj}. We change variables 
as $t_1 \rightarrow u v,\ t_2 \rightarrow u(1-v)$ and perform $v$
integral.   This gives
\begin{equation}
 \Tr [\Psi_{\lambda} Q_{B} \Psi_\lambda   ]
= -\frac{1}{ 2 \pi^2} \int_{0}^{\infty} d u
e^{-u} \left\{
\lambda^2 u^3 + 6 l (1-l) u^2 + 6 (\lambda-1)^2 u
\right\}.\label{235406_17Jun10}  
\end{equation}
At this stage, the integrand still depends on $\lambda $.  However,
it turns out that the $\lambda $ dependence disappears 
once we preform $u$ integration.  More precisely, with the help of the
formula
\begin{equation}
 \int_{0}^{\infty} d u
e^{-u} u^n = \Gamma(n+1) = n!,
\end{equation}
 (\ref{235406_17Jun10}) is evaluated as
\begin{align}
\Tr [\Psi_\lambda Q_{B} \Psi_\lambda   ] & = -\frac{1}{2 \pi^2}
\left(
6 \lambda^2 + 12 \lambda (1- \lambda ) + 6 (\lambda -1)^2 
\right) \notag \\
& = -\frac{3}{\pi^2},
\end{align}
which corresponds to the D-brane tension as
\begin{equation}
 E =  \frac{1}{6 } \Tr [\Psi_{\lambda} Q_{B} \Psi_\lambda   ] 
=  -\frac{1}{2 \pi^2}.
\end{equation}
Our major concern is the $\lambda \rightarrow 0$ limit
in which the solution approaches to the identity based configuration.
Let us see whether singularity occurs in each step of our calculation.
First, we find  negative powers of $\lambda$ in (\ref{234157_17Jun10}) 
which diverges in vanishing $\lambda $ limit.
Therefore $\lambda $ should be kept finite
at this stage.  Once the trace is evaluated explicitly, inverse of
$\lambda$ disappears so we can take the limit.   In fact, 
setting $\lambda$ to zero in
(\ref{235406_17Jun10}) gives correct answer.  This nicely explains
why taking $\lambda \rightarrow 0$ limit before evaluation of
the trace yields singular answer.  

Another gauge invariant quantity, the closed string tadpole,
can be evaluated in a similar way as in  \cite{Erler:2009uj}.
Again, the second term in (\ref{225510_17Jun10})
does not contribute to the tadpole
due to the BRST invariance of the tadpole.  
Then we would like to evaluate
\begin{equation}
 \Tr [ V \Psi_{\lambda}] = \Tr 
\left[ V c \frac{1+ (\lambda -1)K}{1+\lambda K }
   \right],\label{002056_21Jul10}
\end{equation}
where $V= c \bar{c} \mathcal{V}_{matter}$
 is a closed string vertex operator insertion 
at open string midpoint.   The first term of (\ref{002056_21Jul10}) is
evaluated as
\begin{equation}
\Tr 
\left[ V c \frac{1}{1+\lambda K }
   \right]=
\Tr [V c \Omega ] \times 
 \int_{0}^{\infty} dt\, 
e^{-t} (t \lambda)
\end{equation}
While the second term is given by
\begin{align}
\Tr \left[ V c \frac{(\lambda -1)K}{1+\lambda K }
   \right]
& = 
 \frac{1-\lambda}{\lambda} \int_{0}^{\infty} dt\,  
e^{-t}  
 \partial_{t}     \Tr [V c  \Omega^{\lambda t} ] \notag \\ 
& = \frac{1-\lambda}{\lambda} 
\int_{0}^{\infty} dt\,  
e^{-t}  
 \partial_{t}     \Tr [ (t \lambda)  V c  \Omega ] \notag \\ 
& = \Tr [V c \Omega ] \times
 \int_{0}^{\infty} dt\,  
e^{-t} (1-\lambda),\label{002854_5Aug10} 
\end{align}
where we perform scale transformation in the second line of
(\ref{002854_5Aug10}).  Then sum of above two terms gives
\begin{equation}
 \Tr [ V \Psi_{\lambda}] = \Tr [V c \Omega ] \times
 \int_{0}^{\infty} dt\,  
e^{-t} (1-\lambda + t \lambda  ). \label{003624_21Jul10} 
\end{equation}
As similar to the case of the classical action, the last integral in
right hand side of (\ref{003624_21Jul10}) does not depends on $\lambda$.
Furthermore, it coincides with an expected answer of closed string
tadpole on the disk \cite{Ellwood:2008jh}. 
\begin{equation}
 \Tr [ V \Psi_{\lambda}] = \Tr [V c \Omega ] \times 1
= \braket{
\mathcal{V}(i \infty) c(0)
}_{C_1}.
\end{equation}

\section{Simpler solution}

It is known that there is more simpler identity based solution 
\cite{Erler:2009uj,Schnabl:2010tb,Zeze:2010jv}.
\begin{align}
 \Psi_{S} & = - c K \notag \\
 & = \Psi_{A} -c. \label{004002_5Aug10} 
\end{align}
Let us examine the gauge transformation $U_{\lambda}$ for this solution.
Applying it to (\ref{004002_5Aug10}) we have
\begin{align}
 \Psi'_{\lambda} & = \Psi_{\lambda} - U_{\lambda} c U_{\lambda
 }^{-1}\notag \\
 & = -c(1+ \lambda K) B c \frac{K}{1+ \lambda K}\label{003420_5Aug10}
\end{align}
It can be seen that the above string field belongs 
to the second class of solutions, i.e., $\Psi_2$ in (\ref{003651_5Aug10}).

Evaluation of gauge invariant quantities is almost similar to that
of Sec. 2, so we only quote results here.   First, the kinetic term 
of SFT action is evaluated as
\begin{align}
 \Tr [\Psi'_{\lambda} Q_B \Psi'_{\lambda} ]
& = -\frac{3}{\pi^2} \int_{0}^{\infty} d u u\, e^{-u} \notag \\
& = -\frac{3}{\pi^2}. 
\end{align}
Surprisingly,  the integrand does not depend on $\lambda$ even before 
preforming $u$ integration!   Similar phenomena 
also occurs for the closed string tadpole.
\begin{align}
 \Tr[V c \frac{-K}{1+\lambda K}] &
= \frac{1}{\lambda} \int_{0}^{\infty} dt e^{-t}\partial_{t} \Tr[ V c \Omega^{\lambda
 t}] \notag
\\
&= \frac{1}{\lambda} \int_{0}^{\infty} dt e^{-t} \partial_{t} \Tr[ \lambda t V c \Omega]
 \notag \\
 & =   \int_{0}^{\infty} dt e^{-t} \partial_{t} \Tr[ t V c \Omega  ]
 \notag \\
& = \Tr[  V c \Omega].
\end{align}

\section{SFT around the solution}

In this section, we give some remarks about the
string field theory around the solutions discussed in this paper.
The new kinetic operator
$Q'_B$ for the background field $\Psi_\lambda $,
\begin{equation}
 Q'_B \Psi = Q_B \Psi + \Psi_\lambda \Psi + \Psi \Psi_\lambda,
\end{equation}
characterize the spectrum in this background.
First, the homotopy operator for 
(\ref{225510_17Jun10}) is given by
\begin{equation}
 A_{\lambda} = \frac{B}{1+ \lambda K}.
\end{equation}
It can be easily checked that it satisfies
\begin{equation}
Q'_B(A_{\lambda}) =  Q_B  A_{\lambda} +\Psi_{\lambda} A_{\lambda} +
  A_{\lambda}\Psi_{\lambda} = 1.
\end{equation}
In particular, the homotopy operator for the identity based solution 
is simply given by
\begin{equation}
 A_{\lambda =0} = B.
\end{equation}
This gives very simple prescription of gauge fixing.  The existence 
of homotopy operator tells us that
\begin{equation}
Q_{B}' \Psi =0 \rightarrow \Psi =Q_{B}'( B \Psi).\label{000407_18Jun10}
\end{equation}
This means that $\Psi $ is always exact if $B \Psi \neq 0$.  On the
other hand, when $B\Psi =0$ holds, (\ref{000407_18Jun10}) implies
\begin{equation}
Q_{B}' \Psi =0 \rightarrow \Psi = 0.
\end{equation}
Therefore, $B \Psi =0$ completely fixes gauge so as to $\Psi$ being
zero. 

The action of the kinetic operator $Q'_B$ on $K B c$ subalgebra 
is also interesting.  It is given by
\begin{align}
 Q'_B c & = 0, \\
 Q'_B B & = 1, \\
 Q'_B K & = [c, K] (1-K).
\end{align}
Transformation of $c$ implies that it loses the role of infinitesimal vector
of conformal transformation.  $Q'_B B =1$, which already appeared as
the cohomology operator, implies that it also does not  gives a generator
of conformal transformation $K$.  At first look, these two
transformation seems to imply $Q'_{B} \sim c$, which is a reminiscent of
the vacuum string field theory.  However, things are not so simple 
since the third equation is nontrivial with respect to $K$.

Algebra for the simplest identity based solution $\Psi = -c K$ is
more interesting.  It is given by
\begin{align}
 Q'_B c & = 0, \\
 Q'_B B & = 0, \\
 Q'_B K & = [K, c] K.
\end{align}
$Q'_B$ vanishes for $B$ and $c$ both. 

\section{Discussion}

In this paper, we investigate
an one-parameter family of classical solutions in SFT
which interpolates between identity based solution and 
the Erler-Schnabl solution.  Classical action does not depend on
the gauge parameter and also gives correct value of D-brane tension. The
closed string tadpole is also confirmed to give expected answer. 
Since the solution can be made arbitrary close to the 
identity based one, it can be regarded as a consistent 
regularization of an identity based
solution.  To our knowledge, this is the first example of regularization
which correctly reproduce the D-brane tension.  

It is very valuable to discuss why our regularization 
works well and earlier attempts fail. Cleary, a key
feature of our regularizaiton is gauge invariance.  
In gauge theory, gauge invariant regularization 
plays crucial role in evaluation of physical quantities.   
Since our regularization is realized by a gauge
transformation, the equation of motion is ensured and the value of
classical action is kept unchanged as long as the gauge transformation
is regular.  One the contrary, earlier attempts to attatch worldsheets
with small width to a solution violate gauge invarinace thus yield
indefinite result.

It is also interesting to notice that the final form of
the gauge invariant quantity is given by
\begin{equation}
 \int_{0}^{\infty} d u  e^{-u} 
 f (\lambda, u),\label{011741_21Jul10}
\end{equation}
where $f(\lambda, u)$ is a polynomial such that the $\lambda$ dependence
disappears after $u$ integration.  
It should be stressed that the $u$ integration runs
thorough all width even in $\lambda\rightarrow 0$ limit.  
In other words, a gauge invariant contraction 
between identity based string fields 
can be calculated as an correlator on
world sheet with non zero width! 
This fact is trivial from the point of view of gauge invariance,
but also tells us that a reason why a naive regularization of attaching
infinitesimal piece of world sheet has been failed in past.  
It is also expected that other gauge invariant quantity, which is not yet known, takes form of
(\ref{011741_21Jul10}).  Therefore it will be interesting to investigate
possible form of $f(\lambda, u)$ to classify gauge invariant observable.

Existence of consistent regularization of
an identity based solution will play important role 
in feature developments in string field theory.    
In particular, it will be very useful to explore the physics around
close string vacuum since the description of the theory becomes much
simpler.  It will also be useful to find other solutions which are not
gauge equivalent to the Erler-Schnabl solution.   

\section*{Acknowledgment}

We thank H. Isono for valueable discussions and comments.

\end{document}